\documentclass[12pt,preprint]{aastex}
\usepackage{natbib,graphicx}
\bibpunct{(}{)}{;}{a}{}{,}

\slugcomment{Nuclear polarization of FRI radio galaxies}

\shorttitle{Nuclear polarization of FRI radio galaxies}
\shortauthors{Kharb, Shastri, Gabuzda}

\begin{document}

\title{When less is more: Are radio galaxies below the Fanaroff-Riley break 
more polarized on parsec scales~?}

\author{P. Kharb\altaffilmark{1,2}}

\author{P. Shastri\altaffilmark{1}}

\and

\author{D. C. Gabuzda\altaffilmark{3}}

\altaffiltext{1}{Indian Institute of Astrophysics, Bangalore 560034, India.}
\altaffiltext{2}{Presently at the Rochester Institute of Technology, Rochester,
NY 14623, USA. Email: kharb@cis.rit.edu}
\altaffiltext{3}{University College Cork, Cork, Ireland.}

\begin{abstract}
We present images showing the first detections of polarization on parsec 
scales in the nuclei of four Fanaroff-Riley type~I (low-luminosity) radio 
galaxies. Observations with Very Long Baseline Interferometry at 
$\lambda$~3.6cm reveal 
the presence of ordered magnetic fields within $\sim$1650 Schwarzchild radii 
of the putative central supermassive black hole. The relatively high fractional 
polarization in the pc-scale jets of these galaxies is consistent with the 
standard scheme unifying low-luminosity radio galaxies with BL~Lac objects. 
This result also suggests that these radio galaxies lack the obscuring tori 
that apparently depolarize the nuclear emission in the more powerful FRII 
radio galaxies, and that their supermassive blackholes are poorly fed and/or 
inefficient radiators.
\end{abstract}


\section{INTRODUCTION}

\citet{FanaroffRiley74} were the first to point out
that radio galaxies with 178~MHz luminosities below and above
$2\times10^{25}$~WHz$^{-1}$ display different kpc-scale radio
morphologies: radio galaxies below this divide (FRI) have plume-like
structures, while those above this divide (FRII) have well-collimated jets
terminating in so-called hot spots. Based on comparisons of 
orientation-independent properties, it has been widely argued that FRI
radio galaxies constitute the parent population of the strongly 
Doppler beamed BL~Lac objects, while FRII radio galaxies, particularly 
those with narrow emission lines, 
constitute the parent population of radio-loud 
quasars~\citep[][and refs. therein]{UrryPadovani95}.

Magnetic ($B$) fields are believed to be instrumental in the formation and
collimation of the synchrotron-emitting relativistic jets ejected by
radio-loud active galaxies \citep{BlandfordPayne82}
Ordered pc-scale $B$-fields have been detected in active galaxies with 
highly Doppler beamed jets via polarization-sensitive Very Long Baseline 
Interferometry (VLBI) \citep[][and refs. therein]{Gabuzda94}.
However, direct evidence for such fields is meagre in galaxies whose jets are 
believed to be closer to the plane of the sky (and therefore not
heavily relativistically beamed) 
\citep[Taylor, Hough, \& Venturi 2001;][]{MiddelbergTH}. In 
particular, there have been no detections of pc-scale polarization 
for any radio galaxy below the FR luminosity divide.
In this letter we report the detection of appreciable polarization on parsec 
scales in the nuclei of four FRI 
radio galaxies. We discuss the significance of our results in the context
of the FR divide and the standard radio-loud Unification Scheme.

\section{OBSERVATIONS AND DATA REDUCTION}

We observed the four FRI radio galaxies $viz.,$ 3C~66B, 3C~78, 3C~264
and 3C~270, at 8.4 GHz ($\lambda$ 3.6cm) with  
the Very Large Baseline Array 
(VLBA)\footnote[4]{The NRAO is operated for the National Science Foundation 
(NSF) by Associated Universities, Inc. (AUI), under a cooperative agreement.
The European VLBI Network (EVN) is a joint facility of
European, Chinese, South African and other radio astronomy institutes funded
by their national research councils.}
and five stations of the European VLBI Network$^4$
$viz.,$ Effelsberg, Onsala, Yebes, 
Medicina, and the phased WSRT, on 2002 March 1. The observations were made 
in a ``snap-shot'' mode with the observing time per target 
totalling five to eight hours. The total data rate was 128~Mbits/sec, 
with 8~MHz recorded in each of the four baseband converters (BBCs) for each 
right- and left-circular polarization and with two-bit sampling. All antennas
observed both right- and left-circular polarization, except for Onsala and
Yebes, which observed only RCP. The data were correlated at 
the MkIV Data Processor at JIVE.

The calibration and imaging of the data were done using standard techniques
in the AIPS package, using  system temperatures and gain curves provided by 
the NRAO and EVN, and Los Alamos as the reference antenna. 
The unpolarized source 3C~84 was used to determine the instrumental 
polarizations (D-terms) of the antennas using the AIPS task LPCAL. 
The instrumental polarizations for the WSRT were very high ($\sim20\%$); and 
it was not possible to adequately calibrate and remove
these D-terms. We therefore flagged all the WSRT polarization data
before running LPCAL, which led to refined values for the 
remaining D-terms.  The compact, polarized source 1156+295 was used 
to calibrate the polarization position angles, by comparing the polarization
in our VLBI map with the integrated 8.4~GHz polarization measured on 
2002 March 8 from the NRAO VLA  
monitoring database\footnote[5]{http://www.vla.nrao.edu/astro/calib/polar}.

The AIPS tasks IMAGR and CALIB were used to make the total intensity ($I$) 
images. Since Onsala and Yebes recorded only RCP, the polarization
$u-v$ coverage was not symmetric, so that the polarization ($P$) beam was
complex. Accordingly, the AIPS procedure CXPOLN and task CXCLN were
initially used to make the polarization maps. The resulting $P$ images 
had appreciably higher noise levels than those obtained by excluding the 
polarization data for Onsala and Yebes, and we adopted the latter images 
as our final polarization images.
We used a large empty region covering typically $>$ 200 beam areas to
estimate the rms noise of the $I$ image and $\sim$100~beam areas for the 
rms noise of the $P$ image (Table~1).

The spatial resolution of the images is $\sim$0.5~mas, corresponding
to $\sim$0.2--0.4~pc. Significant polarization was detected from the 
inner parsec, $i.e.,$ the ``cores'' and/or inner jets of all four FRI radio 
galaxies. Figure~1 shows the total intensity images of the four FRIs with the 
polarization electric vectors and the distribution of the degree of 
polarization (in color) superimposed.

\section{DISCUSSION}
\label{bgeom}

{\bf Pc-scale Polarization -- $B$-field Geometry}:
Comparisons with VLBI maps with similar resolution obtained at other 
frequencies \citep[e.g.,][]{Giovannini01,JonesWehrle97} confirm that, in
each case, the brightest radio component has a flat or inverted spectral
index, and can be identified as the VLBI ``core.'' The fractional polarization 
in the pc-scale radio cores is $m_{c}\simeq0.4-1\%$, with the degree 
of polarization rising to $m_{i-j}\simeq5-10\%$ in the inner jet 
($\le$~1 pc from the core; see Table~1). We also detected jet polarization 
of $m_{j}\sim$20\% further from the cores of 3C~78 and 3C~264, and 
$m_{j}$ reaches 60\% in a knot about 1.5~pc from the core of 3C~264. 
These values are comparable to those observed for the pc-scale jets of 
BL~Lacs and quasars, indicating the presence of appreciably ordered 
$B$-fields and little depolarization. 

When the source is optically thin to the emitted
synchrotron radiation (as is the case for the jets), the observed
polarization angle is orthogonal to the $B$-field in the emission region,
while the polarization angle and source $B$-field are aligned when the
emission region is optically thick \citep{Pacholczyk70}. In this case, 
the inferred jet $B$-fields within one parsec of the cores of 3C~66B and 3C~78 
are roughly transverse to the local jet direction, while the relative $B$-field 
orientation in the inner jet of 3C~264 is unclear; it may be oblique to the 
flow direction, but no definite conclusion can be drawn because we cannot 
account for possible bends in the jet on scales smaller than our beam or 
for local variations in Faraday rotation.
3C~270 has no detected polarization beyond its core (Fig.~1).

BL~Lac objects most often have transverse jet $B$-fields on parsec scales,
originally taken as evidence for shocks that compress an initially 
disordered field~\citep[e.g.,][]{Gabuzda94}. However, 
Gabuzda, Murray \& Cronin (2004) found that at least some of these jets 
have a transverse rotation-measure gradient, which is a signature of an
ordered toroidal $B$-field component \citep{Blandford93}, possibly associated 
with a helical $B$-field.
The development of an appreciable longitudinal $B$-field component with 
distance from the core has also been observed for a number of BL~Lacs 
\citep[e.g.,][hereafter P05]{Pushkarev05}.
The jet $B$-field geometries in 3C~66B and 3C~78 are both typical
of those that are observed in BL~Lac objects, providing
support for the standard unification picture.
If the transverse field in 3C~66B is associated with an ordered toroidal 
$B$-field component and not shock compression, this component begins to be the 
dominant ordered component on scales as small as $\sim$0.3~pc from the 
core, assuming a black-hole mass of $1.9 \times 10^9$M$_\sun$
(Noel-Storr et al. 2005, preprint) and an inclination
$\theta\simeq$40$\degr$ \citep{Giovannini01}. This corresponds to only
$\sim$1650 Schwarzchild radii from the putative central supermassive
black hole.

Three of the four FRI radio galaxies show one-sided ``core--jet'' 
pc-scale structures as is characteristic of radio-loud active galaxies, 
believed to be due to Doppler boosting/dimming of the 
approaching/receding jet. The fourth, $viz.,$ 3C~270, shows a two-sided jet, 
suggesting that it is inclined closer to the plane of the sky.  
It is noteworthy that the two radio galaxies with the highest 
jet-to-counterjet intensity ratios ($i.e.,$ those expected to be 
most highly Doppler beamed of the four), $viz.,$ 3C~78
and 3C~264, have more detected polarization and structure in their 
jets than the other two sources. 

3C~78 shows a clear alternation of regions with
transverse and longitudinal inferred $B$-field along its jet,
reminiscent of similar alternating $B$-field structures observed in the
BL~Lacs OJ~287 \citep{GabuzdaGomez01} and 1418+546
(P05), which have been interpreted as evidence for
``global'' intrinsic fields associated with the jets.
We also observe a region of roughly longitudinal
$B$-field shifted toward the southern edge of the jet of 3C~264
($\sim1.3$ pc from the core). This could be a consequence of
shearing of the jet due to its interaction with the surrounding medium
\citep{Laing99}, or alternatively may come about because the jets
have a helical $B$-field (Lyutikov, Pariev, \& Gabuzda 2005). Such polarized
``sheaths'' have been observed in the quasar 1055+088 
(Attridge, Roberts, \& Wardle 1999; P05) and a number of BL~Lacs (P05).

{\bf Pc-scale Polarization -- FR Divide and Unification}:
In contrast to our detection of pc-scale polarization in the nuclear
regions of all four of these FRI radio galaxies, such polarization was
detected in only one of four narrow-line FRII radio galaxies
\citep{Taylor01,MiddelbergTH,Zavala02}: weak polarization with $m_{c}
\simeq0.2\%$ was detected only in the flat-spectrum core of 3C~166.
Even if results for broad-line FRII radio galaxies
are included, the core-region polarization detection rate remains only about
25\%: two of eight FRII radio galaxies, $viz.,$ 3C~166 and 3C~111, show
polarization in their 8~GHz pc-scale cores. In contrast, four of eight
FRI radio galaxies show core-region polarization with $m_{c} \ge 0.4\%$.
This detection rate increases to $100\%$ if we exclude
those galaxies that are in cooling flows ($viz.,$ 3C~274, 3C~317, 3C~218)
or recent mergers (NGC~5128), and are therefore expected to be lying in
regions with high Faraday depths \citep{SarazinWise93} that may strongly
depolarize the emission.

The comparatively low core polarization in the FRII radio galaxies may be
due to depolarization by a foreground screen that is fragmented on scales
smaller than the spatial resolution; candidates for such foreground screens
include the inner ionized edge of an obscuring torus \citep{Zavala03},
photoionized clouds in the broad-line and narrow-line regions
\citep{Zavala02}, ionized confining gas in the narrow-line region
\citep{Zavala03} and an accretion disk corona around the central engine
\citep{MiddelbergTH}. While several FRII galaxies show evidence for a
torus \citep{UrryPadovani95}, searches for obscuring tori in FRI galaxies
have not yielded definitive results. NGC~5128 shows signs of a torus-like disk
\citep{Alexander99}, but recent infrared observations of 3C~274 by
\citet{Perlman01} failed to detect emission from a dusty torus. Studies of
FRI galaxies on parsec scales have revealed neutral-hydrogen (HI)
absorption in thin gaseous disks, with the cores being essentially
unobscured \citep{Taylor96,Morganti02}. Our detection of polarized emission
from the core region of 3C~270, which also displays HI and free-free
absorption in a nuclear gas disk, likewise suggests that the circumnuclear
disk is thin (in agreement with the predictions of
\citet{JonesWehrle97} and \citet{Langevelde00}), so that it 
fails to depolarize the emission from the mostly unobscured core.

It is interesting that the core fractional polarization of BL~Lacs and 
quasars (the purported Doppler beamed counterparts of FRIs and FRIIs) 
differ, being $m_{c} \sim2-5\%$ in BL~Lac objects and $m_{c} < 2\%$
in quasars (Gabuzda et al. 1994; Pollack, Taylor, \& Zavala 2003).
The origin of this difference is 
not entirely clear, but space-VLBI polarization observations at 5~GHz have 
shown that the ground-based VLBI ``core'' polarization of BL~Lac objects 
are dominated by the contribution of newly emerging, highly polarized jet 
components \citep[][P05]{Gabuzda99VSOP,GabuzdaGomez01}. The modest core 
polarization in both the BL~Lacs and quasars could be due to the weakness of 
the ordered $B$-field component or depolarization by circumnuclear thermal 
material. \citet{Zavala03} derived smaller rotation measures in the 
pc-scale 
cores of several BL~Lacs than is typical for quasars, and this trend is 
confirmed by results for a much larger sample of BL~Lac objects (Gabuzda
and Pashchenko, in prep.). This suggests that BL~Lac objects have less
ionized gas in their central regions, which could explain both their higher 
core polarization (lower depolarization) and their almost featureless optical 
spectra. This is also consistent with the possibility that FRI sources lack 
both an obscuring torus and substantial amounts of ionized gas, since this 
gas should be unobscured and should therefore give rise to lines in the 
optical spectrum; indeed, Baum, Zirbel, \& O'Dea (1995) have demonstrated that 
the optical emission lines of FRI galaxies are systematically less luminous
than those of FRII galaxies. Note also that the core polarization angles
in 3C~66B and 3C~264 are both nearly transverse to the jet direction;
although we cannot infer the corresponding $B$-field geometry without knowing
whether the region of polarized emission is optically thin or thick, the
small offset of the observed polarization angles from being strictly
orthogonal to the jet suggests an absence of substantial Faraday rotation,
consistent with a dearth of thermal free electrons in the core regions of 
these galaxies.

\section{SUMMARY AND CONCLUSIONS}

Our global VLBI polarimetry of the nuclei of four radio galaxies of the 
FRI type at $\lambda$~3.6cm and a typical angular resolution of 
0.5~mas have yielded the following results:

\noindent
1.~~We have obtained secure detections of polarization between 0.4--1\% 
within a $\sim$parsec sized region in all the nuclei.

\noindent
2.~~Although we are dealing with small numbers, this suggests that the 
detection rate of nuclear polarization for FRI radio galaxies is considerably 
higher than that for FRII galaxies.

\noindent
3.~~If the lower degree of nuclear polarization in FRIIs  
is due to depolarization by the inner ionized edge of an 
obscuring torus, the higher detection rate in the FRIs would suggest 
that the latter lack such a torus or other ionized material around the bases 
of their nuclear jets.

\noindent
4.~~The detected inner-jet polarization is oriented parallel to the 
local direction of the jets in 3C~66B and 3C~78, implying a 
transverse $B$-field in these jets, as is typical of BL~Lac 
objects. In 3C~264 and 3C~270, the orientation of the polarization in the 
inner jet and core, respectively, relative to the local jet direction 
is unclear 
due to possible resolution and/or Faraday rotation effects. We also find 
evidence for the development of a longitudinal $B$-field component further 
from the core in 3C~78 and 3C~264, as well as regions of alternating $B$-field 
geometry in 3C~78, as has been observed for a number of BL~Lac objects.
Taken as a whole, the qualitative and quantitative similarities between the 
total intensity and polarization structures of 3C~66B, 3C~78, and 3C~264 and 
those of BL~Lac objects are striking, and our images 
provide at least tentative evidence in support of the standard unification 
scheme, in which FRIs constitute the parent population of BL~Lac objects. 

\noindent
5.~~We have detected an ordered $B$-field component on scales as small as
$\sim$0.3~pc in 3C~66B. Based on the estimated black-hole mass for 3C~66B
of $\sim2 \times 10^9$M$_\sun$, we conclude that these $B$-fields are located  
within $\sim$1650~Schwarzschild radii of the putative central black hole.

\noindent
6.~~The modest  degree of depolarization inferred for the pc-scale jets 
of these FRIs 
is consistent with a dearth of ionized material. This could be due to 
either the presence of smaller amounts of gas available for ionization, or 
to a lower flux of ionizing photons. The latter, in turn, could be associated 
with either a  
relatively low accretion rate (consistent with the availability
of relatively  
smaller amounts of gas), or alternatively with inefficient 
radiation of the accretion disk \citep{Reynolds96}.  

\acknowledgments
We are thankful to Chris O'Dea for providing the blackhole mass estimate
for 3C~66B, prior to publication.


\begin{table}[h]
\begin{center}
\caption{Measured and derived parameters for the four FRI radio galaxies}
\begin{tabular}{lccccccccccl}
\tableline\tableline 
Source&S$_{VLBI}$&$R_j$&$P_{peak}$&$P_{rms}$&$m_{c}$&$m_{i-j}$&
$m_{o-j}$&$B_{i-j}$& $B_{o-j}$ \\
Name  &(mJy) &&(mJy/bm)&($\mu$Jy/bm)&  \%&\%&\%&  \\
\hline
3C 66B&195 &23.4&0.68&56&1.1 &0.8&...&$\bot$ & ...\\
3C 78 &585 &133.8&1.36&104&...&0.6&4.6&$\bot$ & $\|,\bot$  \\
3C 264&206 &51.4&1.09&76&0.6 &0.7&9.3&Unclear & $\|$\\
3C 270&429 &2.0&0.70&78&0.4 &...&...&...&...  \\
\tableline 
\end{tabular}
\end{center}
Col.~2: Total VLBI flux density in mJy at 8.4 GHz;
Col.~3: Jet-to-counter-jet ratio estimated $\sim$1~pc from the core;
Col.~4: Peak surface brightness of the $P$ map in mJy/beam;
Col.~5: rms noise in the $P$ map in $\mu$Jy/beam;
Col.s~6, 7: Fractional polarization in the VLBI core and
the inner jet within $\sim0.5$~pc of the core, respectively;
Col.~8: Fractional polarization in the outer jet $\sim$0.5 -- 1.0~pc 
from core and
Col.s~9,10: Inferred $B$-field geometry in the inner and
outer jet, respectively, relative to the local jet direction, 
assuming that the jet emission is optically thin
($\bot$ = perpendicular, $\|$ = parallel; Unclear = affected by
resolution and/or Faraday rotation effects, see section \ref{bgeom}). 
\end{table}


\begin{figure}
\centerline{
\includegraphics[height=8.0cm]{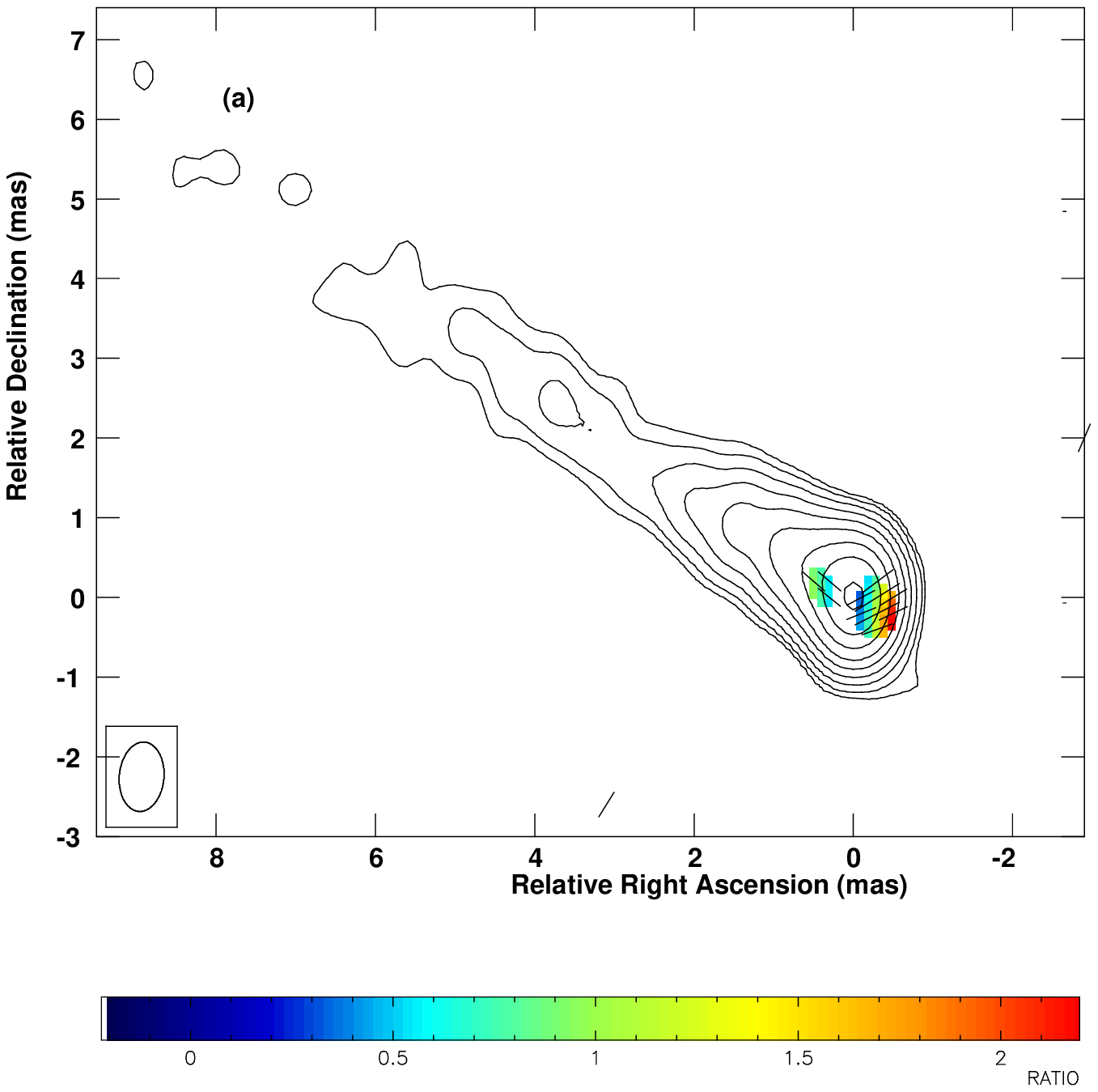}
\includegraphics[height=8.0cm]{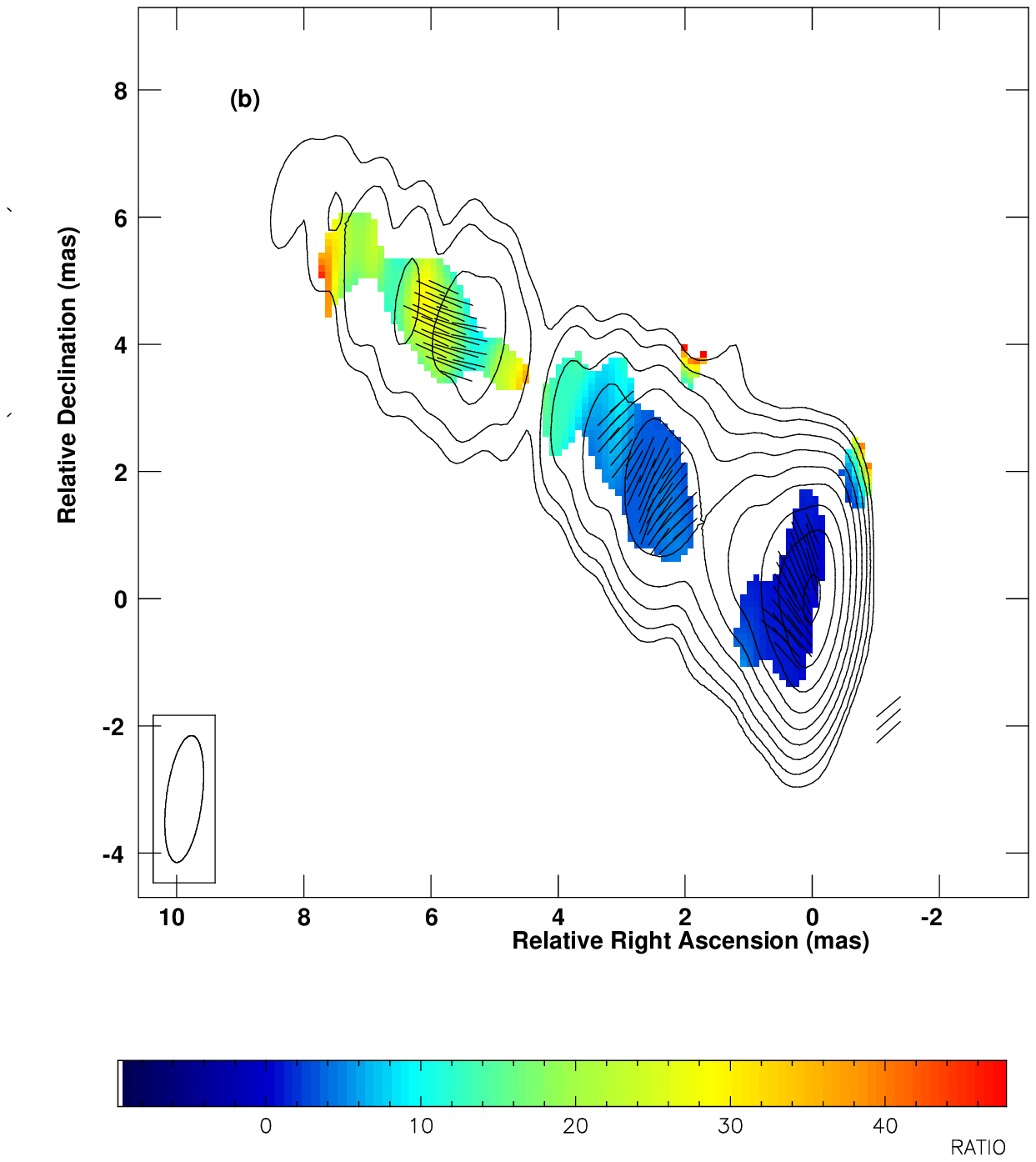}}
\centerline{
\includegraphics[height=8.8cm]{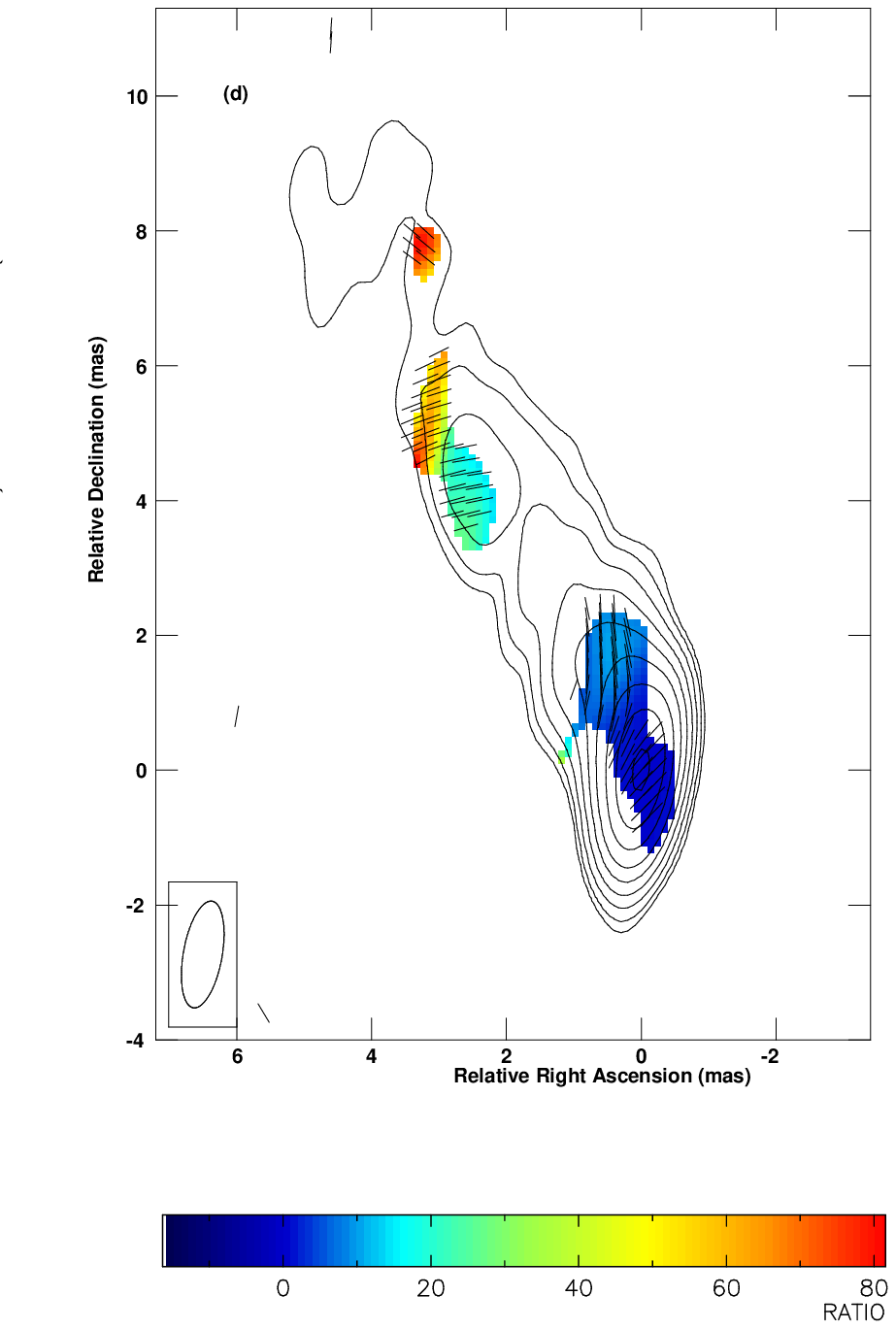}
\includegraphics[height=4.9cm]{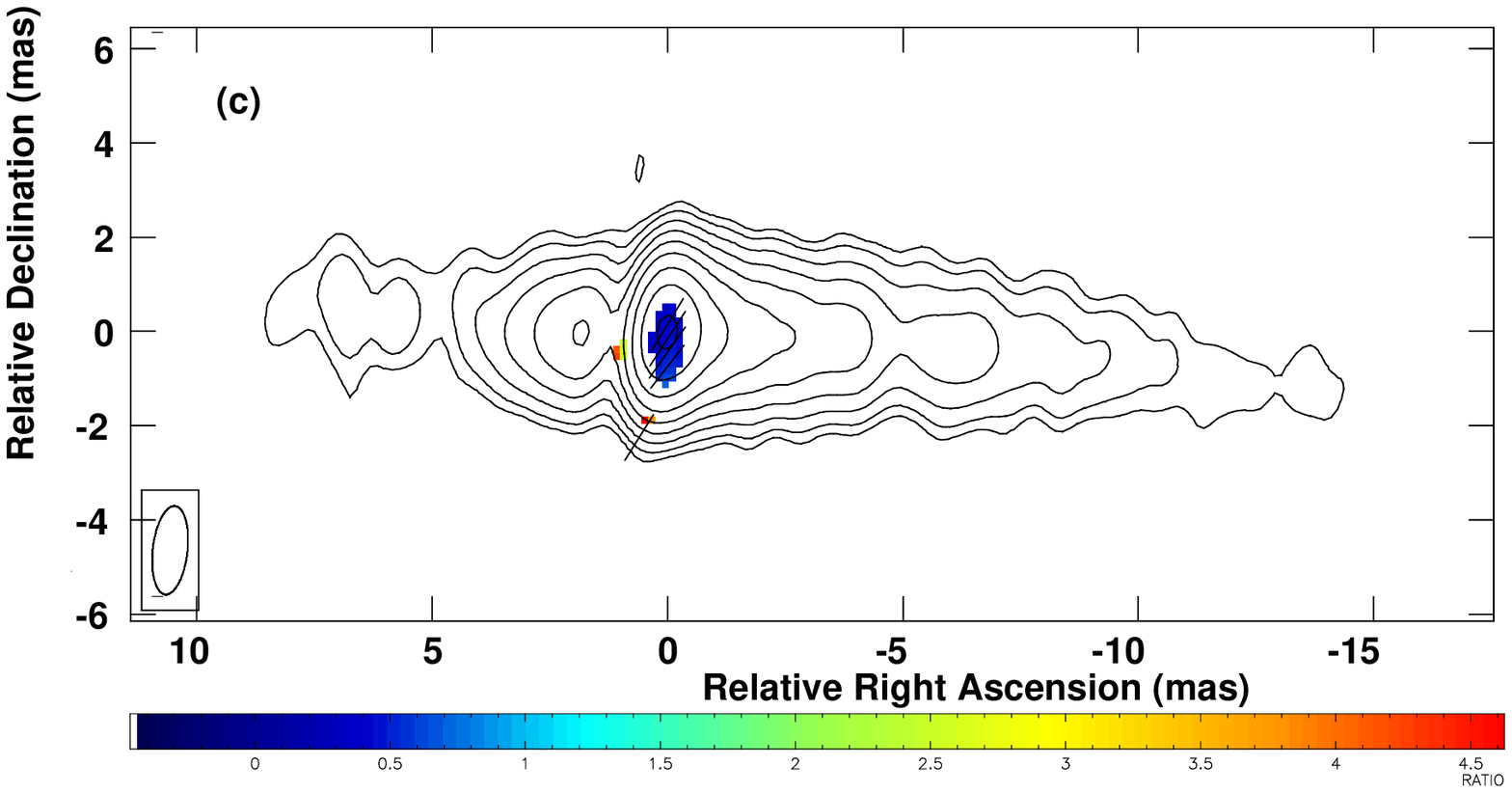}}
\figcaption{
Total intensity maps of four FRI radio galaxies with polarization
electric vectors and the distribution of the degree of polarization 
(in $\%$, in color) superimposed.
{\it Clockwise from top left}: 3C~66B, 3C~78, 3C~270 and 3C~264. 
The surface brightness peaks in mJy/beam are
3C~66B: 118.2, 3C~78: 285.9, 3C~270: 165.3, 3C~264: 136.3.
In all cases the lowest contour is $\pm 0.35\%$ of the peak surface 
brightness, and the contour levels are in percentage of the peak, 
increasing in steps of two.
Both 3C~66B and 3C~78 display predominantly transverse $B$-fields in
their jets, as is the case in BL~Lac objects, while the $B$-field orientation
in the jet of 3C~264 is unclear close to the core 
({\it cf.} section \ref{bgeom}) and predominantly 
longitudinal at larger distances along the jet.}
\end{figure}


\begin{thebibliography}{}

\bibitem[\protect\citeauthoryear{{Alexander}, {Efstathiou}, {Hough}, {Aitken},
  {Lutz}, {Roche} \& {Sturm}}{{Alexander} et~al.}{1999}]{Alexander99}
{Alexander,} D.~M.,  {Efstathiou,} A.,  {Hough,} J.,  {Aitken,} D.,  {Lutz,}
  D.,  {Roche,} P., \&  {Sturm,} E.  1999, \mnras, 310, 78

\bibitem[\protect\citeauthoryear{{Attridge}, {Roberts} \& {Wardle}}{{Attridge}
  et~al.}{1999}]{Attridge99}
{Attridge,} J.~M.,  {Roberts,} D.~H., \&  {Wardle,} J.~C.  1999, \apjl, 518,
  L87

\bibitem[\protect\citeauthoryear{{Baum}, {Zirbel} \& {O'Dea}}{{Baum}
  et~al.}{1995}]{Baum95}
{Baum,} S.~A.,  {Zirbel,} E.~L.,  \&  {O'Dea,} C.~P.  1995, \apj, 451, 88

\bibitem[\protect\citeauthoryear{{Blandford} \& {Payne}}{{Blandford} \&
  {Payne}}{1982}]{BlandfordPayne82}
{Blandford,} R.~D., \& {Payne,} D.~G.  1982, \mnras, 199, 883

\bibitem[\protect\citeauthoryear{{Blandford}}{1993}]{Blandford93}
{Blandford,} R.~D. 1993, {Astrophysical Jets} (Cambridge
University Press)

\bibitem[\protect\citeauthoryear{{Fanaroff} \& {Riley}}{{Fanaroff} \&
  {Riley}}{1974}]{FanaroffRiley74}
{Fanaroff,} B.~L., \& {Riley,} J.~M.  1974, \mnras, 167, 31


\bibitem[\protect\citeauthoryear{{Gabuzda}}{{Gabuzda}}{1999}]{Gabuzda99VSOP}
{Gabuzda,} D.~C.  1999, New Astronomy Review, 43, 691

\bibitem[\protect\citeauthoryear{{Gabuzda} \& {G{\' o}mez}}{{Gabuzda} \& {G{\'
  o}mez}}{2001}]{GabuzdaGomez01}
{Gabuzda,} D.~C., \& {G{\'o}mez,} J.~L.  2001, \mnras, 320, L49

\bibitem[\protect\citeauthoryear{{Gabuzda}, {Mullan}, {Cawthorne}, {Wardle} \&
  {Roberts}}{{Gabuzda} et~al.}{1994}]{Gabuzda94}
{Gabuzda,} D.~C.,  {Mullan,} C.,  {Cawthorne,} T.~V.,  {Wardle,} J.~C.,
\&  {Roberts,} D. 1994, \apj, 435, 140

\bibitem[\protect\citeauthoryear{{Gabuzda}, {Murray} \& {Cronin}}{{Gabuzda}
  et~al.}{2004}]{GabuzdaMurrayCronin04}
{Gabuzda,} D.~C.,  {Murray,} {\' E}.,  \&  {Cronin,} P.  2004, \mnras, 351, L89

\bibitem[\protect\citeauthoryear{{Giovannini}, {Cotton}, {Feretti}, {Lara} \&
  {Venturi}}{{Giovannini} et~al.}{2001}]{Giovannini01}
{Giovannini,} G.,  {Cotton,} W.~D.,  {Feretti,} L.,  {Lara,} L., \&   {Venturi,} T.
  2001, \apj, 552, 508

\bibitem[\protect\citeauthoryear{{Jones} \& {Wehrle}}{{Jones} \&
  {Wehrle}}{1997}]{JonesWehrle97}
{Jones,} D.~L., \& {Wehrle,} A.~E.  1997, \apj, 484, 186

\bibitem[\protect\citeauthoryear{{Laing}, {Parma}, {de Ruiter} \&
  {Fanti}}{{Laing} et~al.}{1999}]{Laing99}
{Laing,} R.~A.,  {Parma,} P.,  {de Ruiter,} H.~R.,  \&  {Fanti,} R.  1999, \mnras,
  306, 513


\bibitem[\protect\citeauthoryear{{Lyutikov}, {Pariev} \& {Gabuzda}}{{Lyutikov}
  et~al.}{2005}]{LyutikovParievGabuzda05}
{Lyutikov,} M.,  {Pariev,} V.~I., \&   {Gabuzda,} D.~C.  2005, \mnras, in press
(astro-ph/0406144)


\bibitem[\protect\citeauthoryear{{Middelberg}}{{Middelberg}}{2004}]{MiddelbergTH}
{Middelberg,} E.  2004, Ph.D. Thesis, Bonn (astro-ph/0404583)


\bibitem[\protect\citeauthoryear{{Morganti}}{{Morganti}
  et~al.}{2002}]{Morganti02}
{Morganti,} R. et~al. 2002, in Proc. of
  the 6th EVN Symp., {Thin disks and HI absorption in the centre of low
  power radio galaxies}, 171

\bibitem[\protect\citeauthoryear{{Noel-Storr}, {Baum}, {Verdoes Kleijn}, {van
  der Marel}, {O'Dea}, {de Zeeuw} \& {Carollo}}{{Noel-Storr}
  et~al.}{2003}]{Noel-Storr03}
{Noel-Storr,} J.,  {Baum,} S.~A.,  {Verdoes Kleijn,} G.,  {van der Marel,} R.,
  {O'Dea,} C.~P.,  {de Zeeuw,} P.~T., \&   {Carollo,} C.  2003, \apjs, 148, 419

\bibitem[\protect\citeauthoryear{{Pacholczyk}}{{Pacholczyk}}{1970}]{Pacholczyk70}
{Pacholczyk,} A.~G.  1970, {Radio astrophysics. Nonthermal processes in
  galactic and extragalactic sources}
(San Francisco: Freeman)

\bibitem[\protect\citeauthoryear{{Perlman}, {Sparks}, {Radomski}, {Packham},
  {Fisher}, {Pi{\~ n}a} \& {Biretta}}{{Perlman} et~al.}{2001}]{Perlman01}
{Perlman,} E.~S.,  {Sparks,} W.~B.,  {Radomski,} J.,  {Packham,} C.,  {Fisher,}
  R.~S.,  {Pi{\~ n}a,} R., \&   {Biretta,} J.~A.  2001, \apjl, 561, L51

\bibitem[\protect\citeauthoryear{{Pollack}, {Taylor} \& {Zavala}}{{Pollack}
  et~al.}{2003}]{Pollack03}
{Pollack,} L.~K.,  {Taylor,} G.~B., \&   {Zavala,} R.~T.  2003, \apj, 589, 733

\bibitem[\protect\citeauthoryear{{Pushkarev}, {Gabuzda}, {Vetukhnovskaya} \&
  {Yakimov}}{{Pushkarev} et~al.}{2005}]{Pushkarev05}
{Pushkarev,} A.~B.,  {Gabuzda,} D.~C.,  {Vetukhnovskaya,} Y.~N., \&   {Yakimov,}
  V.~E. [P05]  2005, \mnras, 356, 859

\bibitem[\protect\citeauthoryear{{Reynolds}, {di Matteo}, {Fabian}, {Hwang} \&
  {Canizares}}{{Reynolds} et~al.}{1996}]{Reynolds96}
{Reynolds,} C.~S.,  {di Matteo,} T.,  {Fabian,} A.~C.,  {Hwang,} U., \&   {Canizares,}
  C.~R.  1996, \mnras, 283, L111

\bibitem[\protect\citeauthoryear{{Sarazin} \& {Wise}}{{Sarazin} \&
  {Wise}}{1993}]{SarazinWise93}
{Sarazin,} C.~L., \& {Wise,} M.~W.  1993, \apj, 411, 55

\bibitem[\protect\citeauthoryear{{Taylor}}{{Taylor}}{1996}]{Taylor96}
{Taylor,} G.~B.  1996, \apj, 470, 394

\bibitem[\protect\citeauthoryear{{Taylor}, {Hough} \& {Venturi}}{{Taylor}
  et~al.}{2001}]{Taylor01}
{Taylor,} G.~B.,  {Hough,} D.~H., \&   {Venturi,} T.  2001, \apj, 559, 703

\bibitem[\protect\citeauthoryear{{Urry} \& {Padovani}}{{Urry} \&
  {Padovani}}{1995}]{UrryPadovani95}
{Urry,} C.~M., \& {Padovani,} P.  1995, \pasp, 107, 803

\bibitem[\protect\citeauthoryear{{van Langevelde}, {Pihlstr{\" o}m}, {Conway},
  {Jaffe} \& {Schilizzi}}{{van Langevelde} et~al.}{2000}]{Langevelde00}
{van Langevelde,} H.~J.,  {Pihlstr{\" o}m,} Y.~M.,  {Conway,} J.~E.,  {Jaffe,} W.,
\&    {Schilizzi,} R.~T.  2000, \aap, 354, L45

\bibitem[\protect\citeauthoryear{{Zavala} \& {Taylor}}{{Zavala} \&
  {Taylor}}{2002}]{Zavala02}
{Zavala,} R.~T., \& {Taylor,} G.~B.  2002, \apjl, 566, L9

\bibitem[\protect\citeauthoryear{{Zavala} \& {Taylor}}{{Zavala} \&
  {Taylor}}{2003}]{Zavala03}
{Zavala,} R.~T., \& {Taylor,} G.~B.  2003, New Astronomy Review, 47, 589

\end{thebibliography}
\end{document}